\documentclass[pre,aps,twocolumn]{revtex4}
\usepackage[pdftex]{graphicx}
\usepackage{subfigure,amsbsy,amsmath}
\usepackage[latin1]{inputenc}

\usepackage{amssymb}
\usepackage{amsfonts}
\usepackage{mathrsfs}

\begin{document}

\title{Connecting curves for dynamical systems}

\author{R. Gilmore$^1$, Jean-Marc Ginoux$^2$, Timothy Jones$^1$, 
C. Letellier$^3$, and  U. S. Freitas$^3$}

\affiliation{$^1$Physics Department, Drexel University, Philadelphia,
  Pennsylvania 19104, USA}

\affiliation{$^2$ UMR 7586 - Institut de Mathematiques de Jussieu, 
Universit\'e Pierre et Marie Curie, Paris VI,}

\affiliation{$^3$ CORIA UMR 6614 - Universit\'e de Rouen, BP 12,
Av. de l'Universit\'e, Saint-Etienne du Rouvray cedex, France}

\date{\today}

\begin{abstract}

We introduce one dimensional sets
to help describe and constrain the integral curves
of an $n$ dimensional dynamical system.
These curves provide more information about the
system than the zero-dimensional 
sets (fixed points) do.  In fact, these curves
pass through the fixed points.  Connecting curves
are introduced using two different but equivalent definitions, one
from dynamical systems theory, the other from differential
geometry.  We describe how to compute these curves and
illustrate their properties by showing the connecting
curves for a number of dynamical systems.\\


\end{abstract}

\pacs{PACS numbers: XXXXXXX}
\maketitle
\section{Introduction}

Poincar\'e proposed that the fixed points
of a dynamical system could be used to provide some information
about, or constraints on, the behavior of trajectories defined
by a set of $n$ nonlinear ordinary differential equations
(a dynamical system) \cite{Poin81, Poin82, Poin85, Poin86}.
The fixed points of a dynamical system
constitute its zero-dimensional invariant set.
Unfortunately, the fixed points provide only local information
about the nature of the flow.

Since that time many, including
Andronov, Tikhonov, Levinson, Wasow, Cole, O'Malley and Fenichel, have
focused on higher dimensional invariant sets, in particular
on $n-1$ dimensional invariant sets.  In many instances these
are slow invariant manifolds of singularly perturbed dynamical
systems.  These manifolds enable one to define the \textit{slow} part
of the evolution of the trajectory curve of such systems.
Until now, it seems that, except for the works of \cite{Roth98},
no one has investigated the problem of \textit{one-dimensional
sets} which play a very important role in the
structure of chaotic attractors by connecting their
fixed points. The aim of this work is to define and
present methods for constructing such
one-dimensional sets.  The sets that we construct are
generally not trajectories that satisfy the equations of the
dynamical system.

The first attempt to study one-dimensional sets
has been made in the context of Fluid Mechanics by Roth and Peikert
\cite{Roth98}. The idea of transporting the concept of ``vortex core
curves'' to the phase space of dynamical systems is due to
one of us (R.G.), who applied it to three-dimensional dynamical
systems and then to higher-dimensional dynamical systems.
In the context of classical differential geometry, another of us (J.M.G.) called such
curves \textit{connecting curves} since they are one-dimensional
sets that connect fixed points.

In Sec. II we set terminology by introducing autonomous
dynamical systems and define the velocity and acceleration
vector fields in terms
of the forcing equations for these systems.
In  Sec. III we introduce the idea of the vortex
core curve through an eigenvalue-like equation
derived from the condition that one of the eigendirections
of the Jacobian of the velocity vector field is
colinear with the velocity vector field.
We show that this defines a one-dimensional curve in
the phase space.  In Sec. IV we introduce the idea
of connecting curves from the viewpoint of differential
geometry.  These are defined by the locus of points where
the curvature along a trajectory vanishes.
In Sec. V we show that the two definitions are equivalent.
In Sec. VI we describe three methods for computing the
connecting curves for a dynamical system.  Several applications
are described in Sec. VII including two models introduced by
R\"ossler, two introduced by Lorenz, and a dynamical
system with a high symmetry.  The figures show clearly that
the connecting curve plays an important role as an axis
around which the flow rotates, which is why this curve is
called the vortex core curve in hydrodynamics.
The figures also underline an observation made in
\cite{Roth98} that this curve is at best an approximation
to the curve around which the flow swirls.
In the final Section we summarize our results and provide
a pointer to visual representations of
many other dynamical systems and their connecting curves.

\section{Dynamical systems}

We consider a system of differential equations defined in a compact $E$
included in $\mathbb{R}^n$ with $\vec {X}=
\left[ {x_1 ,x_2 ,...,x_n } \right]^t\in E\subset \mathbb{R}^n$:

\begin{equation}
\label{eq1}
\frac{d\vec {X}}{dt}= \overrightarrow \Im  ( \vec {X} )
\end{equation}
where $\overrightarrow \Im ( \vec {X} ) =
\left[ f_1 (\vec {X}),f_2 ( \vec {X}),...,f_n ( \vec {X}) \right]^t
\subset \mathbb{R}^n$
defines a velocity vector field in $E$  whose components $f_i$
are assumed to be continuous and infinitely
differentiable with respect to all $x_i $, i.e., are
real-valued $C^\infty $ functions  (or $C^r$ for $r$ 
sufficiently large) in $E$ and which satisfy the
assumptions of the Cauchy-Lipschitz theorem \cite{Codd55}. A solution of this
system is the parameterized \textit{trajectory curve} 
or \textit{integral curve} $\vec {X}\left( t
\right)$ whose values define
the states of the dynamical system described by
Eq. (\ref{eq1}). Since none of the components $f_i$
of the velocity vector field depends here explicitly on time, the system is
said to be \textit{autonomous}.


As the vector function $\vec {X}\left( t \right)$ of the scalar variable $t$
represents the trajectory of a particle M, the total derivative of
$\vec {X}\left( t \right)$ is
the vector function $\overrightarrow V \left( t \right)$ of the scalar
variable $t$ which represents the instantaneous velocity vector of M
at the instant $t$, namely:

\begin{equation}
\label{eq2}
\overrightarrow V \left( t \right)=\frac{d\vec {X}}{dt}=
\overrightarrow \Im ( \vec {X})
\end{equation}
The instantaneous velocity vector $\overrightarrow V \left( t \right)$ is
tangent to the trajectory except at the fixed points, where it is zero.


The time
derivative of $\overrightarrow V \left( t \right)$ is the vector function
$\vec {\gamma }\left( t \right)$ that represents
the instantaneous acceleration vector of M at the instant $t$

\begin{equation}
\label{eq3}
\vec {\gamma }\left( t \right)=\frac{d\overrightarrow V }{dt}
\end{equation}
Since the functions $f_i $ are supposed to be sufficiently differentiable,
the chain rule leads to the
derivative in the sense of Fr\'{e}chet \cite{Frec25}:

\begin{equation}
\label{eq4}
\frac{d\overrightarrow V }{dt}=\frac{\partial \overrightarrow \Im
}{\partial \vec{X}}\frac{d\vec {X}}{dt}
\end{equation}

By noticing that $\frac{\partial \overrightarrow \Im }{\partial \vec {X}}$
is the functional Jacobian matrix $J$ of the dynamical system (\ref{eq1}),
it follows from Eqs. (\ref{eq3}) and (\ref{eq4}) that

\begin{equation}
\label{eq5}
\vec {\gamma }=J\overrightarrow V
\end{equation}
This equation plays  a very important role in the discussions below.

\section{Dynamical Systems and Vortex Core Curves}

At a fixed point in phase space the eigenvectors of the
Jacobian matrix define the local stable and unstable manifolds.
At a general point in phase space the eigenvectors of the
Jacobian with real eigenvalues define natural displacement
directions.  There may be points in the phase space
where two eigenvalues form a complex
conjugate pair and one is real, and the real eigenvector 
is parallel to the vector field that defines the flow.
Under these conditions we expect that the flow in the neighborhood
of such points swirls around the flow direction, much as air flow
swirls around the core of a tornado.  This parallel condition
can be expressed in the coordinate-free form

\begin{equation}
\label{eq9}
J \overrightarrow V  = \lambda \overrightarrow V=\vec {\gamma} 
\end{equation}
The first equation is the mathematical statement of
parallelism; the second equation is a consequence of 
Eq.(\ref{eq5}).

In coordinate form the eigenvalue condition can be written

\begin{equation}
\label{eq8}
\gamma_i = \frac{d}{dt} f_i = \frac{\partial f_i}{\partial
  x_s}\frac{dx_s}{dt} = J_{is}f_s = \lambda f_i~~~~~
1 \le i,s \le 3
\end{equation}
The condition that the acceleration field is proportional
to the velocity field,  $\ddot{x}_i = \lambda \dot{x}_i$
or $\dot{f}_i = \lambda {f}_i$, can be represented
in the form

\begin{equation}
\label{eq6}
\frac{\dot{f}_1}{{f}_1} =\frac{\dot{f}_2}{{f}_2}
=\frac{\dot{f}_3}{{f}_3} =\lambda
\end{equation}
The intersection of the 
surfaces defined by the first two equations defines
a one-dimensional set in the phase space.
This set is a smooth curve that passes
through fixed points.
Alternatively, the three equations define a
one-dimensional set in the phase space
augmented by the eigenvalue $\lambda$:
$(x_1,x_2,x_3,\lambda)$.  The projection of the
one dimensional set from $\mathbb{R}^{3+1}$ down to the phase space $\mathbb{R}^3$
defines the vortex core curve for the dynamical system.

The arguments above are easily extended to define one-dimensional
vortex core curves for $n$-dimensional dynamical systems.

Eq. (\ref{eq9})  has been used to try to identify the
location of the ``vortex core curve'' \cite{Roth98} in
hydrodynamic data.  It is
known that this equation provides a reasonable approximation to the
vortex core when nonlinearities are small but it becomes less useful
as nonlinearities become more important \cite{Roth98}.

\section{Geometry and Connecting Curves}


The approach developed by Ginoux et al. \cite{Gino06, Gino07, Gino09}
uses \textit{Differential Geometry} to study the
metric properties of the trajectory curve, specifically, its
\textit{curvature}\cite{Stru61, Gray06, Krey59}.
A space curve is defined by a set of
coordinates $\vec {X}\left( s \right)$, where $s$
parameterizes the curve.  Typically, $s$ is
taken as the arc length.  When $s$ is instead taken
as a time parameter $t$, derivatives have a natural
interpretation as velocity and acceleration vectors.
The classical curvature along a trajectory 
is defined in terms of the velocity vector  
$\overrightarrow V \left(t \right)$
and acceleration vector 
$\vec {\gamma }\left( t \right)$ by

\begin{equation}
\label{eq10}
\frac{1}{\Re }=\kappa _1 =\frac{\left\| {\vec {\gamma }\wedge
\overrightarrow V } \right\|}{\left\| {\overrightarrow V } \right\|^3}
\end{equation}
Here $\Re $ represents the \textit{radius of curvature}.

We define connecting curves as the curves along which the
curvature $\kappa_1$ is zero.

\textbf{Remark:}  Curvature measures the deviation of
the curve from a straight line in
the neighborhood of any of its points. The location of the
points where the
local curvature of the trajectory curve is
null represents the location of the points of analytical
inflection.

\section{Vortex Core Curves and Connecting Curve}

The dynamical condition Eq.(\ref{eq9}) that defines vortex core curves
can be reexpressed as $ J \overrightarrow V \wedge \overrightarrow V=\vec {0}$.
This is equivalent to the geometric condition Eq.(\ref{eq10}) that
defines connecting curves.  As a result, the two definitions,
one coming from dynamical systems theory, the other
from differential geometry, are equivalent.

Since the two definitions are equivalent, the
conditions they provide for defining the
connecting curve are also identical, as we now show.  The
vanishing conditions for the first curvature of the flow
Eq. (\ref{eq10}) are

\begin{equation}
\label{eq13}
J\overrightarrow V =\lambda \overrightarrow V
\Leftrightarrow
J\overrightarrow V \wedge \overrightarrow V =\vec {0}
\Leftrightarrow
\vec {\gamma }\wedge \overrightarrow V =\vec {0}
\Leftrightarrow
\kappa _1 =0
\end{equation}
By defining: $\phi_{23}=\dot {f_2} f_3 - f_2 \dot {f_3}$, $\phi_{13} = f_1 \dot
{f_3} - \dot {f_1} f_3$ and $\phi_{12}=\dot {f_1} f_2 - f_1 \dot {f_2}$ the third
equality can be rewritten (c.f., Eq.(\ref{eq6}))

\begin{equation}
\label{eq14}
\vec {\gamma }\wedge \overrightarrow V =\vec {0}
\Leftrightarrow
\left\{ {{\begin{array}{*{20}c}
 {\dot {f_2} f_3 - f_2 \dot {f_3} = 0} \hfill \\
 {f_1 \dot{f_3} - \dot {f_1} f_3 = 0} \hfill \\
 {\dot {f_1} f_2 - f_1 \dot {f_2} = 0} \hfill \\
\end{array} }} \right.
\Leftrightarrow
\left\{ {{\begin{array}{*{20}c}
 {\phi_{23}=0} \hfill \\
 {\phi_{13}=0} \hfill \\
 {\phi_{12}=0} \hfill \\
\end{array} }} \right.
\end{equation}

It can be proved that two of the three equations of this nonlinear system
are equivalent and so this relation can be written as three subsystems:

\begin{equation}
\label{eq15}
\left\{ {{\begin{array}{*{20}c}
 {\phi_{23}=0} \hfill \\
 {\phi_{13}=0} \hfill \\
 {\phi_{12}=0} \hfill \\
\end{array} }} \right.
\quad
\Leftrightarrow
\quad
\left\{ {{\begin{array}{*{20}c}
 {\left\{ {{\begin{array}{*{20}c}
 {\phi_{23}=0} \hfill \\
 {\phi_{12}=0} \hfill \\
\end{array} }} \right.} \hfill \\
 {\left\{ {{\begin{array}{*{20}c}
 {\phi_{13}=0} \hfill \\
 {\phi_{12}=0} \hfill \\
\end{array} }} \right.} \hfill \\
 {\left\{ {{\begin{array}{*{20}c}
 {\phi_{23}=0} \hfill \\
 {\phi_{13}=0} \hfill \\
\end{array} }} \right.} \hfill \\
\end{array} }} \right.
\end{equation}

By judiciously choosing one subsystem, say the first, we have another condition
for defining the connecting curve,
i.e. the intersection of two surfaces.

\begin{equation}
\label{eq16}
\left\{ {{\begin{array}{*{20}c}
 {\phi_{23}=0} \hfill \\
 {\phi_{12}=0} \hfill \\
\end{array} }} \right.
\end{equation}

\section{Connecting curve computation}

This kind of problem can not be solved analytically in the general
case. As a result, three numerical approaches have been used to provide
the connecting curve defined by the intersection of
two surfaces, i.e., (\ref{eq16}).

\subsection{First method}

In three dimensions, it is in principle possible to use two of the
three equations $\phi_{ij}(X)=0$ to express two of the
three variables $(x,y,z)$ in terms of the third,
for example $y=y(x),~z=z(x)$.

\subsection{Second method}

If the dynamical system under consideration is of dimension $n$ the
equation $J \vec{V} = \lambda \vec{V}$ represents a set of $n$ equations
in $n +1$ variables: the $n$ coordinates $x_i, i = 1, ..., n$
and the eigenvalue $\lambda$. These $n$ equations define a one-dimensional
set in the enlarged $n + 1$ dimensional space.
The projection of this one-dimensional curve into the $n$
dimensional phase space is the connecting curve
of the dynamical system. Since  $\vec{V} = \vec 0$ at the
fixed points, all fixed points satisfy this equation and thus
belong to the solution set.
The method for constructing the parameterized
version of the connecting curve involves writing down
the $n$ constraint equations for the $n +1$ variables $(x, \lambda)$, and
eliminating all but one.

\subsection{Third method}

As previously observed, the problem for computing the
connecting curve for three-dimensional dynamical systems turns
into the problem of computing the intersection of
two two-dimensional surfaces. A nice method for doing just this
has been developed by Wilkinson \cite{Wilk99}.
We suppose that the intersection of two surfaces
$\phi_{12} \left( {x,y,z} \right)=0$ and
$\phi_{23}\left( {x,y,z} \right)=0$ is
parameterized by $\vec {X}\left( {x\left( t \right),y\left( t \right),z\left(
t \right)} \right)$.
The time derivative of the surface equation leads to
$\nabla \phi_{ij} \left( {x,y,z} \right)\cdot \dot {\vec {X}}=0$. This means
that $\vec {X}$ is perpendicular to both the gradients $\nabla \phi_{ij}
\left( {x,y,z} \right)$ which are the normal vectors to
each surface. As long as these vectors are linearly independent for points
on the intersection, then $\dot {\vec {X}}$ is collinear to the cross
product $\nabla \phi_{12} \left( {x,y,z} \right)\wedge \nabla \phi_{23}
\left( {x,y,z} \right)$

\begin{equation}
\label{eq17}
\dot {\vec {X}}\left( t \right)=\lambda \left( t \right)\nabla \phi_{12}
\left( {x,y,z} \right)\wedge \nabla \phi_{23} \left( {x,y,z} \right)
\end{equation}
By rescaling the time $t$ it is possible to set $\lambda(t)=1$.
Then Eq. (\ref{eq17}) simplifies to the form of an
associated dynamical system (A.D.S.):

\begin{equation}
\label{eq18}
\frac{d\vec {X}\left( t \right)}{dt}=\nabla \phi_{12} \left( {x,y,z}
\right)\wedge \nabla \phi_{23} \left( {x,y,z} \right)
\end{equation}
These equations are generally different from, but related to,
the original dynamical system equations.  The curves defined by this 
equation are not heteroclinic trajectories of the original dynamical system.

Initial conditions for the (A.D.S.)
are any point on the connecting curve,
or any point belonging to the intersection of both surfaces. Thus, the
connecting curve may be defined as the trajectory, or integral, of the
(A.D.S.).  This method is useful as long as the gradients along
the intersection remain nonzero and non-colinear.

\section{Applications}

In this Section we describe the connecting curves
for three- and four-dimensional dynamical systems.

\subsection{R\"{o}ssler model}

The flow equations for the R\"ossler attractor \cite{Ross76} are

\begin{equation}
\label{eq20}
\overrightarrow V \left( {{\begin{array}{*{20}c}
 {\dot{x}} \hfill \\
 {\dot{y}} \hfill \\
 {\dot{z}} \hfill \\
\end{array} }} \right)=\overrightarrow \Im \left( {{\begin{array}{*{20}c}
 {f_1\left( {x,y,z} \right)} \hfill \\
 {f_2\left( {x,y,z} \right)} \hfill \\
 {f_3\left( {x,y,z} \right)} \hfill \\
\end{array} }} \right) = \begin{pmatrix}
 {-y-z} \\
 {x+ay} \\
 {b+z\left( {x-c} \right)} \\
\end{pmatrix}
\end{equation}
where $a$, $b$ and $c$ are real parameters.
The connecting curve for this dynamical system was
computed using all three methods described in Sec. IV.
The solution using the third method has been
performed with Mathematica 7 (files are available at: http://ginoux.univ-tln.fr).

Of the three solution methods just described, the
second leads to the simplest expressions
for the connecting curve.
The curve along which $J\vec{V} = \lambda \vec{V}$ depends on
the three control parameters $(a,b,c)$ and is
parameterized by one of the three phase space coordinates.
Choosing $x$ as the phase space coordinate,
the eigenvalue $\lambda$
satisfies a fifth degree equation

\begin{equation}
\sum_{j=0}^5 D_j \lambda^j = 0
\end{equation}
The coefficients $D_j$
are listed in Table I.  At each fixed point,
the value of $\lambda$ is the value of the real
eigenvalue of the Jacobian matrix at that fixed point.
The coordinates $y$ and $z$ are expressed as rational
functions of $x$ and $\lambda(x;a,b,c)$.  These
rational expressions are

\begin{equation}\begin{array}{c}
y = \displaystyle
\frac{-b-x +ax(c-x) +\lambda x (x-c+a-\lambda)}
{a+(c-x)(1-a^2)+\lambda a(c-x+\lambda -a)} \\
  \\
z = \displaystyle
\frac{+b+x+(\lambda x+ab)(\lambda -a)}
{a+(c-x)(1-a^2)+\lambda a(c-x+\lambda -a)} \end{array}
\end{equation}

\begin{table}
\caption{Coefficients of the fifth degree equation
that defines the eigenvalue $\lambda$ in the expression
for the curve along which the velocity and acceleration
vectors are parallel for the R\"ossler dynamical system.}
\[
\begin{array}{rcl}
D_5 &=& a \\
D_4 &=& 2a(c-a-x) \\
D_3 &=& ax^2-2acx+4a^2x-4a^2c+a^3+c+2a+ac^2\\
D_2 &=&-2a^2x^2+x^2-2a^3x-2cx+4a^2cx-4ax\\
 & & +ab+2ac-2a^2c^2+2a^3c+c^2-2a^2\\
D_1 &=& a^3x^2+4a^2x-2a^3cx-2a^2b+a+b+c-3a^2c+a^3c^2\\
D_0 &=& x^2-a^2x^2+2a^2cx-2cx-2ax+ac-a^2c^2+c^2\\
 & &  -ab+a^3b
\end{array}
\]
\end{table}

\begin{center}
\begin{figure}[t]
\vspace{6mm}
\centerline{\includegraphics[angle=0,width=7cm]{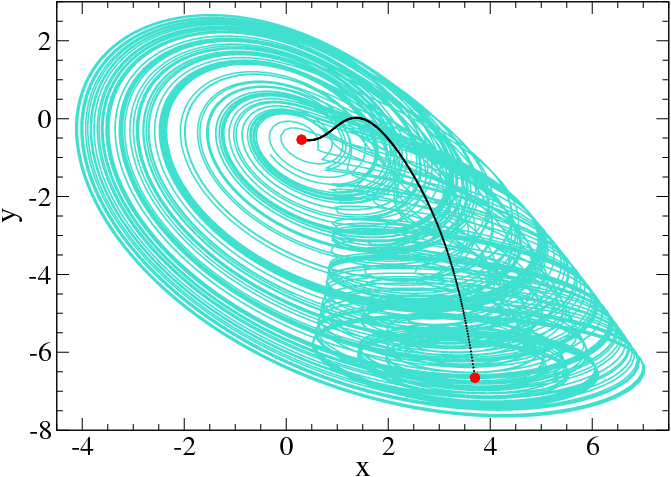}}
\vspace{9mm}
\centerline{\includegraphics[angle=0,width=7cm]{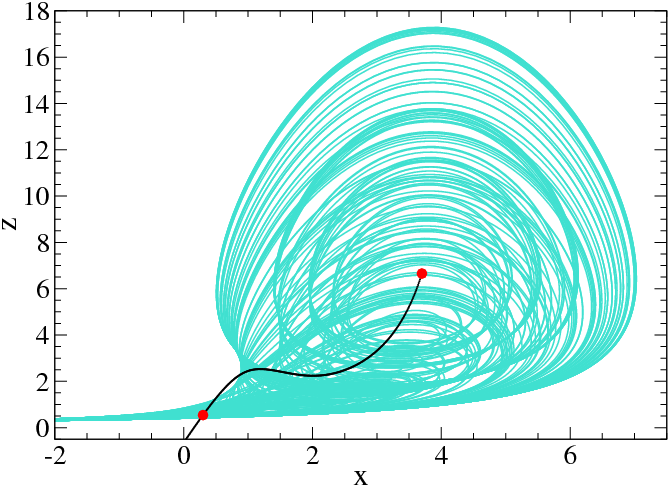}}
\caption{Connecting curve of the R\"ossler model.  The curve
intersects the attractor twice, as seen in the
$x$-$z$ projection.  Parameter values:
$(a,b,c)=(0.556,2,4)$.}
\label{fig1}
\end{figure}
\end{center}

The segment of the connecting curve between
the fixed points (dots) is plotted for the R\"ossler attractor
in Fig. \ref{fig1} for control parameter values
$(a,b,c)=(0.556, 2.0, 4.0)$.  Two projections
are shown.
Near the outer fixed point with repelling real
eigendirection, this curve is a good approximation
to a curve that defines the core of the tornado-like
motion.  However, as it moves toward the fixed
point near the $x$-$y$ plane and the nonlinearities
increase in strength, it becomes a poorer and poorer
approximation of such a curve, even intersecting the
attractor twice before joining the inner fixed point.
This problem is apparent in the $x$-$z$ projection.
This result reinforces an observation made by Roth and Peikert
that the ``eigencurve'', i.e. the
connecting curve defined by $J\vec{V} = \lambda \vec{V}$,
is a good approximation to the vortex core curve in
regions where the nonlinearities are weak, but
not where the nonlinearities become strong \cite{Roth98}.

\subsection{Lorenz model}

The purpose of the model established by Edward Lorenz \cite{Lore63} was
initially to analyze the unpredictable behavior of weather.
After having developed non-linear partial differential equations
starting from the thermal equation and Navier-Stokes equations,
Lorenz truncated them to retain only three modes. The most
widespread form of the Lorenz model is as follows:

\begin{equation}
\label{eq21}
\overrightarrow V \left( {{\begin{array}{*{20}c}
 {\dot{x}} \hfill \\
 {\dot{y}} \hfill \\
 {\dot{z}} \hfill \\
\end{array} }} \right)=\overrightarrow \Im \left( {{\begin{array}{*{20}c}
 {f_1\left( {x,y,z} \right)} \hfill \\
 {f_2\left( {x,y,z} \right)} \hfill \\
 {f_3\left( {x,y,z} \right)} \hfill \\
\end{array} }} \right) = \begin{pmatrix}
  {\sigma \left( {y - x} \right)} \\
 { Rx - y - xz }  \\
 {- b z+xy}
\end{pmatrix}
\end{equation}
where $\sigma $, $R$ and $b$ are real parameters.
Once again, the connecting curves were computed using
all three methods described in Sec. IV.  The calculation
using the third method was performed with
Mathematica 7 (Files are available at: http://ginoux.univ-tln.fr).
All methods gave the same curves.

\begin{center}
\begin{figure}[t]
\vspace{6mm}
\centerline{\includegraphics[angle=0,width=7cm]{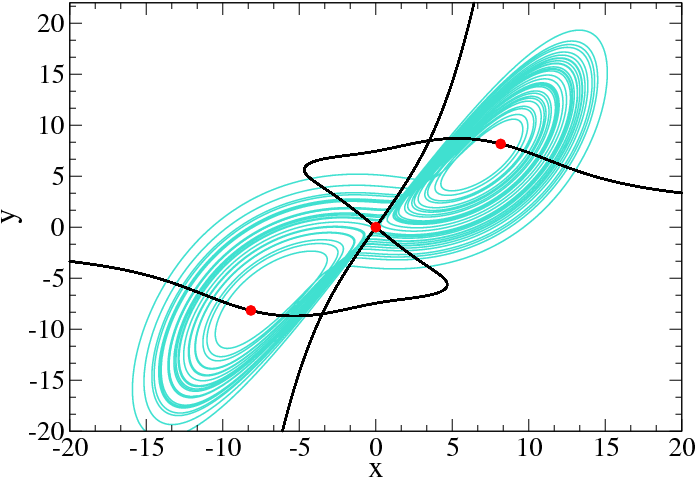}}
\vspace{9mm}
\centerline{\includegraphics[angle=0,width=7cm]{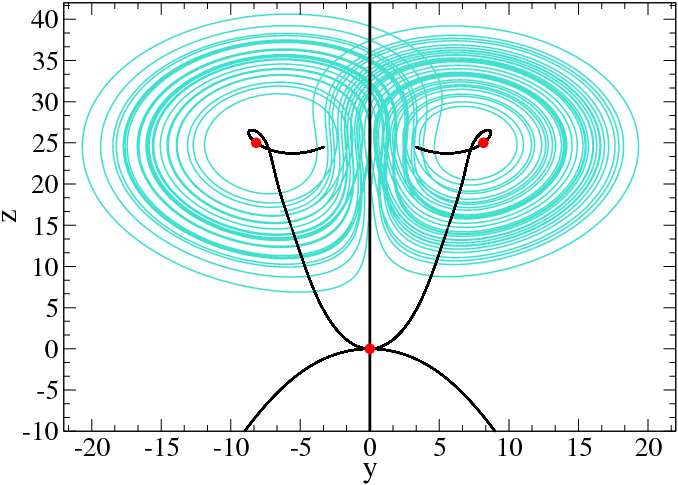}}
\caption{Connecting curve of the Lorenz model.    
One nontrivial connecting
curve heads off to $z \rightarrow -\infty$ and has little effect
on the structure of the attractor.
The other nontrivial connecting curve
connects all three fixed points, and is plotted extending
through the foci.  The third connecting curve is the $z$ axis.
Parameter values:
$(R,\sigma,b) = (28,10,8/3)$.}
\label{fig2}
\end{figure}
\end{center}

Three connecting curves pass through the saddle
at the origin: one corresponding to each of the
three eigendirections with real eigenvalues.
The simplest of
these curves is the $z$-axis, which is simple to compute
by hand.  This particular curve is a trajectory of the Lorenz model.
  A second heads off to $z \rightarrow -\infty$ and
has little effect on the attractor.
The third connecting
curve passes through all three fixed points.  This curve is
shown in Fig. \ref{fig2} in both the $x$-$y$
and $y$-$z$ projections for $(R,\sigma,b)=(28,10,8/3)$.
When $R$ is increased, the return flow from one
side of the attractor to the other exhibits a
fold and the connecting curve intersects
the attractor at the fold.  This reflects a similar
property shown by the R\"ossler equations.

The connecting curves present additional constraints on the structure
of the Lorenz attractor above and beyond those implied
by the location and stability of the fixed points.
Specifically, the flow spirals around and away from the
connecting curve that passes through the two foci.
In addition, the $z$ axis also provides some structure
on this flow, as the flow also always passes in the same direction
around this axis \cite{Byrn04}.

\subsection{Lorenz model of 1984}

In 1984 Lorenz proposed a global atmospheric circulation
model in truncated form \cite{Lore84}.  The model consists of three
ordinary differential equations:

\begin{eqnarray}
\overrightarrow V \left( {{\begin{array}{*{20}c}
 {\dot{x}} \hfill \\
 {\dot{y}} \hfill \\
 {\dot{z}} \hfill \\
\end{array} }} \right) & = & \overrightarrow \Im \left( {{\begin{array}{*{20}c}
 {f_1\left( {x,y,z} \right)} \hfill \\
 {f_2\left( {x,y,z} \right)} \hfill \\
 {f_3\left( {x,y,z} \right)} \hfill \\
\end{array} }} \right) \nonumber \\ 
& = & \begin{pmatrix}
  {-y^2-z^2-a(x-F)} \\
 {-y+xy-bxz+G }  \\
 {bxy+xz-z}
\end{pmatrix}
\end{eqnarray}
In this model the variable $x$ represents the strength
of the globally circling westerly wind current
and also the temperature gradient towards the pole.
Heat is transported poleward by a chain of large scale
eddies.  The strength of this heat transport is
represented by the two variables $x$ and $y$, which
are in quadrature.  The control parameters $aF$ and $G$
represent thermal forcing.  The parameter $b$ describes the
strength of displacement of the eddies by the westerly current.

\begin{center}
\begin{figure}[htbp]
\vspace{6mm}
\centerline{\includegraphics[angle=0,width=7cm]{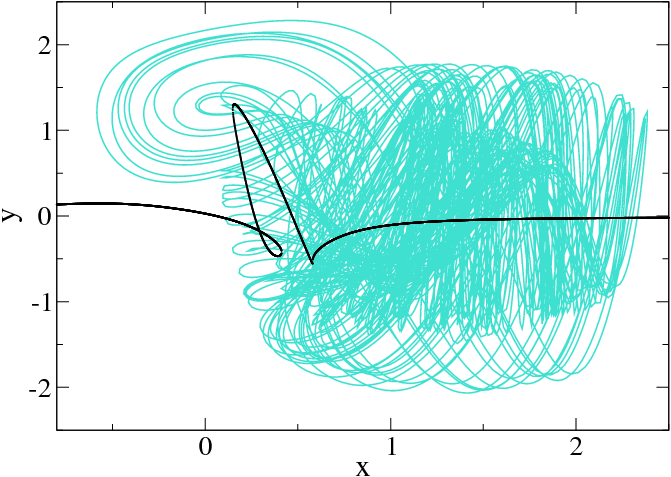}}
\vspace{9mm}
\centerline{\includegraphics[angle=0,width=7cm]{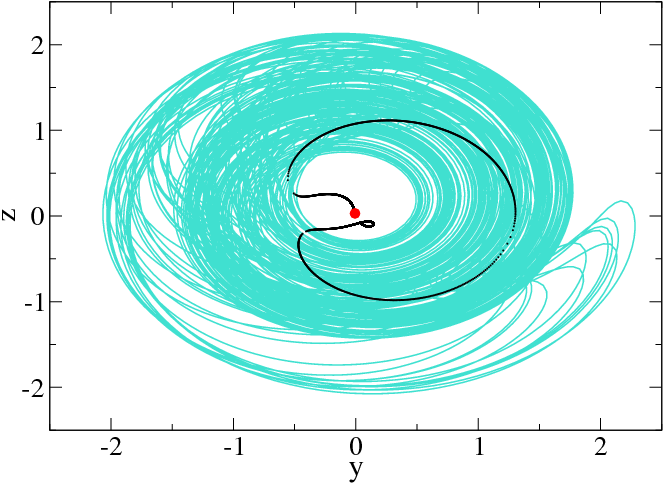}}
\caption{Strange attractor generated by the Lorenz global
circulation model of 1984.  The connecting curve threads
through the inside of the attractor, and is caressed
by the attractor where the stretching and folding is most
pronounced.  Parameter values: $(a,b,F,G) = (1/4,4,8,1)$.}
\label{fig3}
\end{figure}
\end{center}

In Fig. \ref{fig3} we show two projections of this attractor
for control parameters $(a,b,F,G) = (1/4,4,8,1)$ as well
as the connecting curve.  For this set of parameter values
there are three fixed points, only one of which is real at
$(x,y,z)=(7.996,-0.00653,0.0298)$.
It is clear that the connecting
curve goes through the hole in the middle of the attractor,
and that the attractor winds around part of the connecting
curve where most of the bending and folding of the
attractor occurs.  The connecting curve in the $x$-$y$ projection
passes through the fixed point off scale to the right.

\subsection{R\"ossler model of hyperchaos}

\begin{center}
\begin{figure}[htbp]
\vspace{6mm}
\centerline{\includegraphics[angle=0,width=7cm]{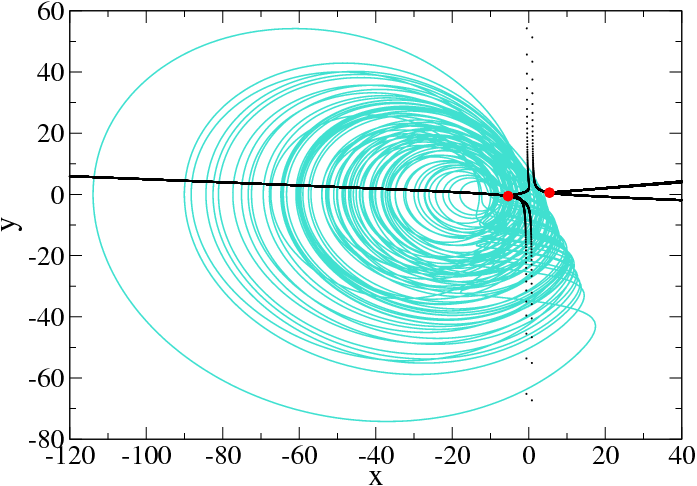}}
\vspace{9mm}
\centerline{\includegraphics[angle=0,width=7cm]{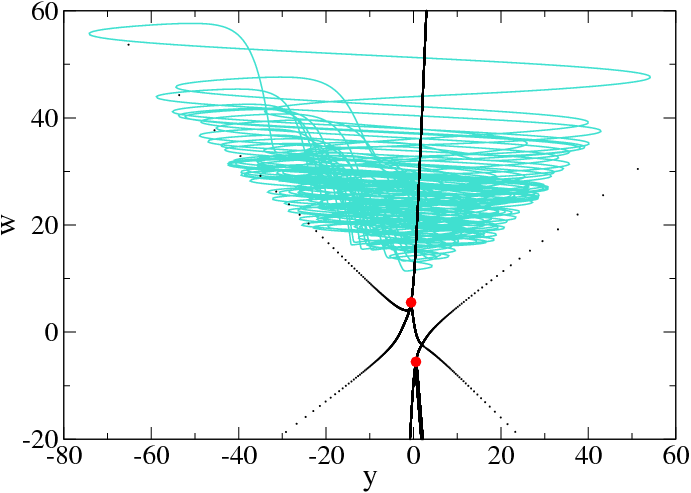}}
\caption{Hyperchaotic attractor generated by the
1979 R\"ossler model for hyperchaos.  Parameter values:
$(a,b,c,d)=(1/4,3,1/2,1/20)$.}
\label{fig4}
\end{figure}
\end{center}

R\"ossler proposed a simple four-dimensional model
in 1979 to study hyperchaotic behavior \cite{Ross79}.  This model is

\begin{eqnarray}
\overrightarrow V \left( {{\begin{array}{*{20}c}
 {\dot{x}} \hfill \\
 {\dot{y}} \hfill \\
 {\dot{z}} \hfill \\
 {\dot{w}} \hfill \\
\end{array} }} \right) & = & \overrightarrow \Im \left( {{\begin{array}{*{20}c}
 {f_1\left( {x,y,z,w} \right)} \hfill \\
 {f_2\left( {x,y,z,w} \right)} \hfill \\
 {f_3\left( {x,y,z,w} \right)} \hfill \\
 {f_4\left( {x,y,z,w} \right)} \hfill \\
\end{array} }} \right) \nonumber \\
& = & \begin{pmatrix}
  {-y-z} \\
 {x+ay+w}  \\
 {b+xz}  \\
 {-cz+dw}
\end{pmatrix}
\end{eqnarray} 
Here the state variables are $(x,y,z,w)$ and the control
parameters are $(a,b,c,d)$.  The connecting curve was computed
using methods 1 and 2 of Sec. IV.  The first method gave
very complicated results.  Method 2 gave simpler results
when the coordinate $z$ was used to express the behavior
of the remaining four variables.  The eigenvalue $\lambda$
was expressed as the root of a seventh degree polynomial
equation whose coefficients were functions of the
four control parameters $(a,b,c,d)$ and $z$.  The
remaining three coordinates were rational functions
of small degree in the variables $z$ and $\lambda(z;a,b,c,d)$.
Two projections of the hyperchaotic attractor and
the connecting curve are shown in Fig. \ref{fig4}.  The computation
was carried out for $(a,b,c,d)=(1/4,3,1/2,1/20)$.  
The fixed points are shown as large dots along the connecting curve.
It is clear from this figure that the connecting curve
provides information about the structure of the attractor,
as the flow in the attractor swirls around the connecting
curve.

\subsection{Thomas Model}

\begin{center}
\begin{figure}[htbp]
\centerline{\includegraphics[angle=0,width=7cm]{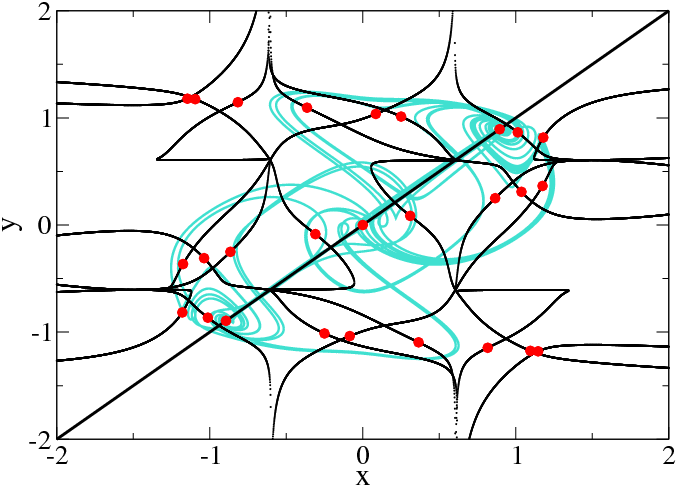}}
\vspace{9mm}
\centerline{\includegraphics[angle=0,width=7cm]{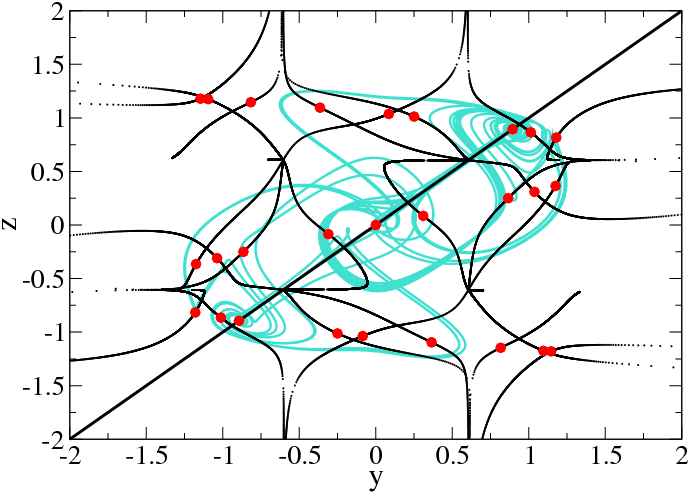}}
\caption{Connecting curves for the Thomas attractor.
One connecting curve is the symmetry axis $x=y=z$.
The remaining connecting curves exhibit the six-fold
symmetry of the system and seek out the holes
in the attractor.  Parameter values:
$(a,b)=(1.1,0.3)$.}
\label{fig5}
\end{figure}
\end{center}

Thomas proposed the following model of a feedback circuit
with a high degree of symmetry \cite{Thom99}:

\begin{eqnarray}
\overrightarrow V \left( {{\begin{array}{*{20}c}
 {\dot{x}} \hfill \\
 {\dot{y}} \hfill \\
 {\dot{z}} \hfill \\
\end{array} }} \right) & = & \overrightarrow \Im \left( {{\begin{array}{*{20}c}
 {f_1\left( {x,y,z} \right)} \hfill \\
 {f_2\left( {x,y,z} \right)} \hfill \\
 {f_3\left( {x,y,z} \right)} \hfill \\
\end{array} }} \right) \nonumber \\
& = & \begin{pmatrix}
  {-bx+ay-y^3} \\
 {-by+az-z^3 }  \\
 {-bz+ax-x^3 }
\end{pmatrix}
\end{eqnarray}
This set of equations exhibits the six-fold rotation-reflection
symmetry $S_6$ about the $(1,1,1)$ axis.  The symmetry
generator is a rotation about this axis by $2\pi/6$ radians
followed by a reflection in the plane perpendicular to
the axis.  The origin is always a fixed point
and, for $a-b>0$, there are two on-axis fixed
points at $x=y=z=\pm \sqrt{b-a}$.  For
$(a,b)=(1.1,0.3)$ there are 24 additional off-axis
fixed points.  These fall into four sets of symmetry-related
fixed points (sextuplets).  One point in each sextuplet is
$(0.085,1.037,0.309),$
$(0.250,1.013, 0.865),$
$(0.364, -1.095, 1.175),$
$(1.146,-1.180,-0.816)$.
The remaining points in a multiplet
are obtained by cyclic permutation
of these coordinates: $(u,v,w)\rightarrow
(w,u,v) \rightarrow (v,w,u)$ and inversion
in the origin $(u,v,w) \rightarrow (-u,-v,-w)$.
The chaotic attractor for this dynamical system
is shown in Fig. \ref{fig5}, along with the
symmetry-related connecting curves and the
27 fixed points.  One of the connecting curves
is the rotation axis.   This is an invariant set
that connects the three on-axis fixed points.
It therefore cannot intersect the attractor.
In fact, this set has the same properties as the
$z$-axis does for the Lorenz attractor of 1963
\cite{Byrn04}.  The remaining connecting curves
trace out the holes in the attractor.  In this sense
they provide additional constraints on the
structure of the attractor over and above those
provided by the spectrum of fixed points.

\section{Discussion}

In this work we go beyond the zero-dimensional invariant
sets (fixed points) that serve to a limited extent to
define the structure of an attracting set of a dynamical system.
We have introduced a curve that we call a connecting curve,
since it passes through fixed points of an autonomous
dynamical system.  We have defined this curve in two
different ways: dynamically and kinematically.
It is defined a vortex core curve dynamically through an eigenvalue-like equation
$J \vec{V} = \lambda \vec{V}$, where $\vec{V}(x)$ is the velocity vector field defining
the dynamical system and $J_{ij}=\partial V_i/\partial x_j$ is
its Jacobian.  We have defined a connecting curve kinematically as the locus
of points in the phase space where the principal curvature
is zero.  These two definitions are equivalent.

Three methods were introduced for constructing this curve
for autonomous dynamical systems.  They were applied
to the standard R\"ossler and Lorenz attractors, where
their behavior with respect to the attractors is shown
in Figs. 1 and 2.  In the figures shown for these attractors,
it is clear that the flow rotates around the connecting curves,
which therefore help to define the structure of the attractor.
The connecting curves were also
constructed for a later Lorenz model, the global atmospheric
circulation model of 1984, and for a later model introduced by
R\"ossler to study chaotic behavior in four dimensional
phase spaces.    Finally, a multiplicity of connecting curves
was computed for an attractor with a high degree of symmetry,
the Thomas attractor.  This is shown in Fig. 5.  
The flows shown in Figs. 3, 4, and 5 are clearly organized by
their connecting curves.
In this sense the connecting curve provide additional
important information about the structure of an attractor,
over and above that provided by the number, nature, and
distribution of the fixed points.
A number of other connecting curves
have been computed, and can be seen at http://www.physics.drexel.edu/\verb+~+tim/programs/.

\section*{Acknowledgements}  

This work is supported in part
by the U.S. National Science Foundation under grant
PHY-0754081.  R. G. thanks CORIA for an invited position.


\end{document}